\documentclass[
pra,twocolumn,showpacs,preprintnumbers,amsmath,amssymb]{revtex4}

\usepackage{graphicx}
\usepackage{amssymb}
\usepackage{amsmath}
\usepackage{dsfont}
\usepackage{bm}
\usepackage{mathrsfs}
\usepackage{times}

\begin{document}
\title{Lipkin's conservation law, Noether's theorem, and the relation to optical helicity}

\author{T.\ G.\ Philbin}

\email{t.g.philbin@exeter.ac.uk}

\affiliation{Physics and Astronomy Department, University of Exeter,
Stocker Road, Exeter EX4 4QL, United Kingdom}

\begin{abstract}
A simple conserved quantity for electromagnetic fields in vacuum was discovered by Lipkin in 1964. In recent years this ``zilch" has been used as a measure of the chirality of light. The conservation of optical zilch is here derived from a simple symmetry of the standard electromagnetic action. The symmetry transformation allows the identification of circularly polarized plane waves as zilch eigenstates. The same symmetry is present for electromagnetism in a homogeneous, dispersive medium, allowing the derivation of the zilch density and flux in such a medium. Optical helicity density and flux are also derived for a  homogeneous, dispersive medium. For monochromatic beams in vacuum, optical zilch is proportional to optical helicity. This monochromatic zilch-helicity relation acquires a factor of the square of the phase index in a dispersive medium.
\end{abstract}

\maketitle

\section{Introduction}
Field theories such as electromagnetism possess an infinite number of conserved quantities in the absence of sources~\cite{lip64,mor64,oco64,kib65,can65,anc05}. One can view electromagnetic conservation laws as constraints on the field dynamics and, in this sense, every conserved quantity is physically significant: the list of conserved quantities is an alternative statement of information contained in Maxwell's equations. On the other hand, only the information contained in a limited number of conserved quantities has a straightforward relation to basic physical concepts. In 1964 Lipkin~\cite{lip64} discovered an electromagnetic conserved quantity that is as simple as those related to the basic concepts of energy, momentum and angular momentum. The density and flux of Lipkin's ``zilch", as he termed it, are
\begin{align} 
\rho_\chi=&\frac{\varepsilon_0}{2}\left(-\bm{E\cdot}\partial_t\bm{B}+\bm{B\cdot}\partial_t\bm{E}\right),   \label{rhochi}   \\
\bm{S}_\chi=&\frac{\varepsilon_0}{2}\bm{E\times}\partial_t\bm{E}+\frac{1}{2\mu_0}\bm{B\times}\partial_t\bm{B}.  \label{Schi} 
\end{align}
These obey the conservation law
\begin{equation}  \label{con}
\partial_t\rho_\chi+\bm{\nabla\cdot S}_\chi=0,
\end{equation}
showing that the integral of the density $\rho_\chi$ over all space is conserved. In recent years $\rho_\chi$ has been used as a measure of the chirality density of light, giving a simple method of exploring light configurations that couple to material chirality~\cite{tan10,hen10,yan11,tan11,hen12,sch12,ros12}. One example of a light beam with a non-zero zilch density $\rho_\chi$ is a circularly polarized plane wave, and this has motivated researchers to compare and contrast zilch with the spin angular momentum of light and the helicity of the photon~\cite{afa96,bli11,and12,col12,bar12,cam12}. It is important to note that there are more interesting possibilities than circular polarization for non-zero values of $\rho_\chi$, once local variations are considered. In particular, it has been shown that the evanescent fields of metal patterns can be engineered to produce an enhanced local zilch density~\cite{hen10,hen12,sch12} and light beams with orbital angular momentum can have a locally enhanced $\rho_\chi$ on the axis of the beam~\cite{ros12}. 

The conservation of zilch must be associated with a symmetry of the electromagnetic action
\begin{equation} \label{act}
\mathcal{S}[\bm{A},\phi]=\frac{\varepsilon_0}{2}\int d^4x \left(\bm{E\cdot E}-c^2\bm{B\cdot B}\right),
\end{equation}
which is a functional of the vector potential $\bm{A}(\bm{r},t)$ and the scalar potential $\phi(\bm{r},t)$ with
\begin{equation} \label{pots}
\bm{E}=-\bm{\nabla}\phi-\partial_t\bm{A}, \quad \bm{B}=\bm{\nabla\times A}.
\end{equation}
Once a symmetry of an action is found, Noether's theorem provides a constructive proof of the associated conservation law (see~\cite{wei}, for example). If however a conservation law is discerned using the field equations, as was the case with Lipkin's discovery~\cite{lip64}, then the identification of the associated symmetry of the action may not be straightforward. The relation of zilch to a symmetry transformation has been considered several times before~\cite{cal65,des76,sud86,kri89,prz94,ibr09,cam12b,bli12}, but apart from~\cite{cal65} the transformations considered were not of the scalar and vector potential in the standard electromagnetic action (\ref{act}). Instead actions with additional dynamical variables were considered~\cite{des76,sud86,kri89,prz94,ibr09,cam12b,bli12}, for example functionals of the vector potential, scalar potential and the electric field, functionals of the transverse vector potential and the electric field, or functionals of the vector potential and an ``electric" vector potential. In some cases the particular non-standard action is solely motivated by the desire to derive the conservation law from an action as easily as possible~\cite{des76,sud86,kri89,prz94,ibr09}. In other cases the form of the non-standard action is motivated by an interest in deriving a conservation law associated with electric-magnetic duality~\cite{cam12b,bli12}, the conserved quantity in this case being optical helicity. Incorporating electric-magnetic duality into the standard action presents some difficulties~\cite{cam12b,bli12}, as  can be seen in~\cite{des76} where the relevant symmetry transformation contains an inverse Laplacian operator (and the action was also rewritten as a functional of the transverse vector potential and the electric field). Calkin's brief note~\cite{cal65} does not directly derive zilch conservation from a transformation of the dynamical variables $\bm{A}$ and $\phi$ of the standard action (\ref{act}), as discussed in Sec~\ref{sec:vac}. 

Our goal here is to derive zilch conservation from a symmetry of the standard action (\ref{act}) using Noether's theorem. As the action (\ref{act}) is a functional of the independent degrees of freedom of the electromagnetic field up to gauge invariance, the symmetry underlying zilch conservation takes its simplest form in terms of these dynamical variables (Sec.~\ref{sec:vac}). The symmetry is used in Sec.~\ref{sec:eigen} to construct zilch eigenstates, in a manner analogous to energy, momentum and angular momentum eigenstates. Section~\ref{sec:gen} outlines generalizations of the symmetry transformation that lead to an infinite class of conserved quantities~\cite{lip64,mor64,oco64,kib65,can65,anc05}. In Sec.~\ref{sec:med} we consider the electromagnetic action in a homogeneous, dispersive medium; the zilch symmetry is present for this action and we use it to find the zilch density and flux in such a medium. The relation of zilch and helicity to the notion of optical chirality is still debated~\cite{tan10,hen10,yan11,tan11,hen12,sch12,ros12,afa96,bli11,and12,col12,bar12,cam12}. In Sec.~\ref{sec:hel} we generalize the expressions for optical helicity density and flux to a homogeneous, dispersive medium and show that zilch, helicity and spin angular momentum all have a different dependence on the material dielectric functions.

\section{Optical zilch in vacuum}   \label{sec:vac}

It is easy to show that the vacuum electromagnetic action (\ref{act}) is invariant (up to surface terms) under the following active transformation of the dynamical variables $\bm{A}(\bm{r},t)$ and $\phi(\bm{r},t)$:
\begin{equation}   \label{trans}
\delta\bm{A}=\eta\bm{\nabla\times}\partial_t\bm{A}, \qquad \delta\phi=0,
\end{equation}
where $\eta$ is the infinitesimal parameter of the transformation. Two unusual features of the transformation (\ref{trans}) compared to the Poincar\'{e}-group symmetries of (\ref{act}) are the absence of any change in $\phi$ and the occurrence of two derivatives in the infinitesimal transformation of $\bm{A}$. If either of the derivatives in $\delta\bm{A}$ are removed the result is not a symmetry transformation, but any even number of additional space and time derivatives can be inserted to give new symmetry transformations (see Sec.~\ref{sec:gen}). Calkin~\cite{cal65} considered transformations of the form $\delta\bm{A}=\eta\bm{\nabla\times}\bm{Z}$ for any $\bm{Z}$ satisfying $\bm{\nabla\cdot Z}=0$. Note that such a transformation does not include (\ref{trans}) because the three components of $\bm{A}$ are independent in the action (\ref{act}). Calkin's approach allows a quick extraction of optical zilch from the action (\ref{act}), but here we seek the conservation law from a symmetry transformation of the dynamical variables of the action, as is done for energy-momentum and angular momentum~\cite{wei}.

Noether's theorem~\cite{wei} shows that if we let the parameter $\eta$ in the symmetry transformation (\ref{trans}) depend on space and time, i.e.\ $\eta=\eta(\mathbf{r},t)$, then the change in the action (\ref{act}) under (\ref{trans}) can be written in the form
\begin{equation} \label{actvar}
\delta\mathcal{S}=\int d^4x\left(\rho_\chi\partial_t\eta+\bm{S}_\chi\bm{\cdot\nabla}\eta\right),
\end{equation}
where $\rho_\chi$ and $\bm{S}_\chi$ obey the conservation law (\ref{con}). We now show that $\rho_\chi$ and $\bm{S}_\chi$ in (\ref{actvar}) are optical zilch density and flux, respectively. After various integrations by parts (surface terms are dropped) the form (\ref{actvar}) is achieved with
\begin{align}
\rho_\chi=&\varepsilon_0\left(-\bm{E\cdot}\partial_t\bm{B}-\frac{1}{2}\partial_t\bm{B\cdot}\partial_t\bm{A}\right)-\frac{1}{2\mu_0}\bm{A\cdot}\nabla^2\bm{B},  \label{rho1}  \\
\bm{S}_\chi=&\varepsilon_0\left(-\phi\partial_t^2\bm{B}+\frac{1}{2}\partial_t\bm{A\times}\partial_t^2\bm{A}\right)          \nonumber \\
&-\frac{1}{\mu_0}\left[\partial_t\bm{B\times}\bm{B}-(\bm{\nabla\cdot}\partial_t\bm{A})\bm{B}-\frac{1}{2}\bm{A\times}\nabla^2\partial_t\bm{A}   \right.   \nonumber \\
& \left.\qquad\quad -\frac{1}{2}\partial_tA_i\bm{\nabla}B^i+\frac{1}{2}B^i\partial_t\bm{\nabla}A_i\right].   \label{S1}
\end{align}
We use tensor notation and the summation convention, so that all index summations are contractions of an upper and a lower index. One can verify that (\ref{rho1}) and (\ref{S1}) satisfy the conservation law (\ref{con}) because of the dynamical equations of the action (\ref{act}), namely Maxwell's equations 
\begin{equation} 
\bm{\nabla\cdot E}=0,  \qquad
\bm{\nabla\times B}=\frac{1}{c^2}\partial_t\bm{E}. \label{max}
\end{equation}
(The other two Maxwell equations are identities because of (\ref{pots}).)

The conserved quantities that emerge directly from Noether's theorem applied to the action (\ref{act}) are not gauge invariant~\cite{wei}. Gauge-invariant densities and fluxes are found by the addition of terms that identically satisfy the conservation law. As regards densities such as $\rho_\chi$, the divergence of any field quantity $\bm{X}(\bm{A},\phi)$ can be added without changing the value of the conserved quantity $\int d^3\bm{r}\,\rho_\chi$; if the time derivative of $\bm{X}(\bm{A},\phi)$ is then subtracted from $\bm{S}_\chi$, the conservation law (\ref{con}) is preserved. One verifies that the addition to (\ref{rho1}) of the divergence
\begin{equation}  \label{quant}
\nabla_i\left(-\frac{\varepsilon_0}{2}\phi\partial_tB^i+\frac{1}{\mu_0}A_j\nabla^{[i}B^{j]}\right),
\end{equation}
where the square brackets denote anti-symmetrization (i.e.\ $T^{[ij]}=(T^{ij}-T^{ji})/2$), produces the zilch density (\ref{rhochi}). To preserve the conservation law the time derivative of the quantity in brackets in (\ref{quant}) must be subtracted from (\ref{S1}); this does not produce a gauge-invariant flux but any term with zero divergence can now be added to $\bm{S}_\chi$ while maintaining (\ref{con}). The simplest course is to show that, after the subtraction just described, the divergence of (\ref{S1}) is equal to the divergence of (\ref{Schi}) (without use of Maxwell's equations (\ref{max})). This demonstrates that  (\ref{rhochi}) and (\ref{Schi}) are gauge-invariant versions of (\ref{rho1}) and (\ref{S1}), respectively.

The symmetry transformation (\ref{trans}) is simpler than transformations that have previously been associated with the conservation of optical zilch~\cite{cal65,des76,sud86,kri89,prz94,ibr09,cam12b,bli12}. This is because (\ref{trans}) is a symmetry of the standard electromagnetic action (\ref{act}) that is a functional of the independent degrees of freedom of electromagnetism up to gauge transformations. The standard quantization of electromagnetism is based on (\ref{act}) and it is most natural to relate conservation of optical zilch to a symmetry of this action, as is done for conservation of energy-momentum and angular momentum~\cite{wei}.

\section{Optical zilch eigenstates}   \label{sec:eigen}
The infinitesimal versions of continuous symmetry transformations reveal the generators of the symmetry group associated with a conserved quantity. Eigenstates of these generators have eigenvalues that completely determine the value of the conserved quantity carried by the eigenstate. This is most familiar in quantum mechanics but the same procedure can be applied to classical field theories. The generators $\{i\partial_t,i\bm{\nabla}\}$ of space-time  translation symmetry give plane waves $e^{i\bm{k\cdot r}-i\omega t}$ as energy-momentum eigenstates. The components of the vector generator $\bm{L}=i\bm{r\times \nabla}$ of rotations do not commute but eigenstates of $L^2$ can be found and these give angular-momentum eigenstates~\cite{jac}. The eigenstates are taken to be complex, with the electromagnetic fields given by the real part of the eigenstates. From (\ref{trans}) the generator of the symmetry associated with optical zilch is $i\partial_t\bm{\nabla\times}$ and it acts on the vector potential. We find the vector-potential zilch eigenstates by considering the Fourier-transformed field $\bm{\tilde{A}}(\bm{r},\omega)$ in the frequency domain; note that this field is complex for real time-domain fields $\bm{A}(\bm{r},t)$, so we do not need to consider complex $\bm{A}(\bm{r},t)$ to find the eigenstates. The zilch eigenvalue equation for  $\bm{\tilde{A}}(\bm{r},\omega)$ is then
\begin{equation}  \label{eigenA}
\bm{\nabla\times}\bm{\tilde{A}}(\bm{r},\omega)=\frac{\chi}{\omega}\bm{\tilde{A}}(\bm{r},\omega),
\end{equation}
where $\chi$ is the eigenvalue. We consider a gauge in which $\phi=0$. Then 
\begin{equation} \label{EA}
\bm{\tilde{E}}(\bm{r},\omega)=i\omega\bm{\tilde{A}}(\bm{r},\omega)
\end{equation}
and (\ref{eigenA}) gives
\begin{equation}   \label{eigenBE}
\bm{\tilde{B}}(\bm{r},\omega)=-i\frac{\chi}{\omega^2}\bm{\tilde{E}}(\bm{r},\omega).
\end{equation}
Equation~(\ref{eigenA}) states directly that $\bm{\tilde{B}}(\bm{r},\omega)=(\chi/\omega)\bm{\tilde{A}}(\bm{r},\omega)$; taking the curl of this and using (\ref{eigenA}) gives
\begin{align}
\bm{\nabla\times \tilde{B}}(\bm{r},\omega)&=\frac{\chi}{\omega}\bm{\nabla\times \tilde{A}}(\bm{r},\omega)=\frac{\chi^2}{\omega^2}\bm{\tilde{A}}(\bm{r},\omega)   \nonumber \\
&= -i \frac{\chi^2}{\omega^3}\bm{\tilde{E}}(\bm{r},\omega),  \label{eigenmax}
\end{align}
where the last step uses (\ref{EA}).
Comparing (\ref{eigenmax}) with the second Maxwell equation (\ref{max}) in the frequency domain, we see that the eigenvalue $\chi$ must be
\begin{equation}  \label{chival}
\chi=\pm\frac{\omega^2}{c}.
\end{equation}
Insertion of (\ref{chival}) in (\ref{eigenBE}) gives
\begin{equation}   \label{eigenBE2}
\bm{\tilde{B}}(\bm{r},\omega)=\mp \frac{i}{c}\bm{\tilde{E}}(\bm{r},\omega),
\end{equation}
which is precisely the relation between the complex amplitudes of the electric and magnetic fields in a circularly polarized plane wave~\cite{jac}. We did not assume that the zilch eigenstates are monochromatic---no restrictions were placed on the frequency-domain spectrum $\bm{\tilde{A}}(\bm{r},\omega)$. We see from (\ref{chival}) however that $\chi$ depends on frequency, so for it to be an eigenvalue of the total field in the time domain we can only have one frequency component. Moreover (\ref{eigenBE2}) shows that this monochromatic wave must be circularly polarized. Circularly polarized plane waves are thus zilch eigenstates with  eigenvalues (\ref{chival}). Note that there is degeneracy in the eigenstates: only the frequency, not the direction, of the circularly polarized plane wave determines the eigenvalue (\ref{chival}). Any superposition of plane waves with the same frequency and circular polarization is also a zilch eigenstate with eigenvalue (\ref{chival}). 

If we replace the zilch generator with its quantum version, namely $i\hbar\partial_t\bm{\nabla\times}$, the eigenvalues (\ref{chival}) become
\begin{equation}  \label{chivalq}
\chi=\pm\frac{\hbar\omega^2}{c}.
\end{equation}
The eigenvalues (\ref{chivalq}) per photon energy $\hbar\omega$ are then $\pm\omega/c$, which is well known to be the time-averaged zilch per unit energy of a classical circularly polarized plane wave~\cite{tan10,hen10,yan11,tan11,hen12,sch12,ros12}. 

It is not surprising that the zilch eigenstates should distribute the zilch density uniformly throughout the wave, which is what the circularly polarized plane wave does. For applications however, beams with nonuniform, locally enhanced zilch density are more interesting~\cite{hen10,hen12,sch12,ros12}.

In Sec.~\ref{sec:med} the vacuum zilch eigenvalues (\ref{chival}) and the vacuum time-averaged results for the zilch of plane waves will be generalized to the case of a dispersive medium.

\section{Generalizations of the symmetry transformation}   \label{sec:gen}
As noted in Sec.~\ref{sec:vac},  the transformation (\ref{trans}) is still a symmetry of the action (\ref{act}) when any even number of space and time derivatives are inserted into the transformation of $\bm{A}$. The total number of extra derivatives inserted must be even, so an odd number of space derivatives together with an odd number of time derivatives is allowed. Depending on the number of extra space derivatives and whether the associated indices have any contractions, we can construct an infinite number of conserved quantities with any number of tensor indices. The resulting infinite class of conservation laws were found by Morgan~\cite{mor64}. As an example we note that the generalized symmetries that give a infinite class of {\it scalar} conserved quantities are 
\begin{equation}   \label{transgen}
\delta\bm{A}=\eta\bm{\nabla\times}\nabla^{2n}\partial^{2m}_t\bm{A}, \qquad \delta\phi=0,
\end{equation}
where $n$ and $m$ are any non-negative integers. The conservation law associated with (\ref{transgen}) is calculated as in the $n=m=0$ case. After some tedious manipulations to bring the density and flux into their simplest gauge-invariant forms, we find the conservation law
\begin{equation}  \label{connm}
\partial_t\rho_{(n,m)}+\bm{\nabla\cdot S}_{(n,m)}=0,
\end{equation}
with
\begin{align} 
\rho_{(n,m)}=&\frac{\varepsilon_0}{2}\left(-\bm{E\cdot}\nabla^{2n}\partial^{2m+1}_t\bm{B}+\bm{B\cdot}\nabla^{2n}\partial^{2m+1}_t\bm{E}\right),   \label{rhonm}   \\
\bm{S}_{(n,m)}=&\frac{\varepsilon_0}{2}\bm{E\times}\nabla^{2n}\partial^{2m+1}_t\bm{E}+\frac{1}{2\mu_0}\bm{B\times}\nabla^{2n}\partial^{2m+1}_t\bm{B}.  \label{Snm} 
\end{align}
It is straightforward to verify that (\ref{rhonm}) and (\ref{Snm}) satisfy (\ref{connm}) with use of Maxwell's equations (\ref{max}). The Laplacian operators $\nabla^2$ in (\ref{rhonm}) and (\ref{Snm}) can be replaced by the operator $(1/c^2)\partial^{2}_t$ because of the wave equation.

\section{Optical zilch in a dispersive medium}   \label{sec:med}
Electromagnetic waves in materials can carry conserved quantities in limited frequency ranges where absorption is negligible. The effects of dispersion however can be important even when absorption can be ignored. In this regime of negligible absorption the best known example of a conserved quantity in a dispersive medium is the energy of a monochromatic wave, whose time-averaged energy density is given by Brillouin's formula (\ref{brill}) below (see~\cite{jac}, for example). The electromagnetic energy density has a more complicated expression for waves that are not monochromatic~\cite{phi11}. The electromagnetic momentum density~\cite{phi11} and angular momentum density~\cite{phi12} that measure the conserved values of these quantities inside the medium also have dispersive contributions, which take a simple form in the monochromatic case. The conserved electromagnetic energy-momentum and angular momentum in a dispersive medium can be calculated using Noether's theorem~\cite{phi11,phi12}, and we now do the same for electromagnetic zilch.

The dielectric functions in a limited frequency range where absorption is negligible can be fitted to an even series in frequency (see~\cite{agr}, for example):
\begin{equation} \label{matseries}
\varepsilon(\omega)=\sum_{n=0}^\infty\varepsilon_{2n}\omega^{2n}, \quad \kappa(\omega)=\sum_{n=0}^\infty \kappa_{2n}\omega^{2n},
\end{equation}
where $\varepsilon(\omega)$ is the relative permittivity and the relative permeability is $\mu(\omega)=\kappa(\omega)^{-1}$. The series in (\ref{matseries}) are a fit to the dispersion data of the material in the frequency range of interest and as such will be finite; we do not place an upper limit on the summations as the results will be valid for infinite series. We assume the medium is homogeneous as this will be a requirement for conservation of optical zilch. The $\bm{D}$ and $\bm{H}$ fields in the frequency and time domains are
\begin{gather*}
\bm{\tilde{D}}(\bm{r},\omega)=\varepsilon_0\varepsilon(\omega)\bm{\tilde{E}}(\bm{r},\omega), \  \bm{D}(\bm{r},t)=\varepsilon_0\varepsilon(i\partial_t)\bm{E}(\bm{r},t),    \\
 \bm{\tilde{H}}(\bm{r},\omega)=\kappa_0\kappa(\omega)\bm{\tilde{B}}(\bm{r},\omega),   \ \bm{H}(\bm{r},t)=\kappa_0\kappa(i\partial_t)\bm{B}(\bm{r},t),
\end{gather*}
where $\kappa_0=\mu_0^{-1}$. The electromagnetic action in the homogeneous dispersive medium (\ref{matseries}) is~\cite{phi11,phi12}
\begin{equation} \label{actmed}
\mathcal{S}[\bm{A},\phi]= \! \int \! d^4x\frac{\kappa_0}{2}\left\{\frac{1}{c^2}\bm{E}\cdot[\varepsilon(i\partial_t)\bm{E}]-\bm{B}\cdot[\kappa(i\partial_t)\bm{B}]\right\}.
\end{equation}
Variation of  $\phi$ and $\bm{A}$ in (\ref{actmed}) gives the macroscopic Maxwell equations
\begin{gather} 
\varepsilon_0\bm{\nabla\cdot}[\varepsilon(i\partial_t)\bm{E}]=0,  \label{macmax1}  \\
\kappa_0\bm{\nabla\times}[\kappa(i\partial_t)\bm{B}]=\varepsilon_0\varepsilon(i\partial_t)\partial_t\bm{E}. \label{macmax2}
\end{gather}

The action (\ref{actmed}) is still symmetric under the transformation (\ref{trans}) (this transformation is not a symmetry if the medium is inhomogeneous, i.e.\ if $\varepsilon(i\partial_t)\to\varepsilon(\bm{r},i\partial_t)$ and/or $\kappa(i\partial_t)\to\kappa(\bm{r},i\partial_t)$). The conservation law associated with the symmetry (\ref{trans}) is extracted from the action (\ref{actmed}) using Noether's theorem, as in the vacuum case of Sec.~\ref{sec:vac}. There is now the extra complication of the  series in (\ref{actmed}). The method of dealing with these series when applying Noether's theorem is described in~\cite{phi11} and~\cite{phi12}, to which we refer the reader for the details. The zilch density and flux in the resulting conservation law can be made gauge invariant as in Sec.~\ref{sec:vac} and the result is
\begin{align}
\rho_\chi=&-\frac{1}{2}\bm{D\cdot}\partial_t\bm{B}+\frac{1}{2}\bm{B\cdot}\partial_t\bm{D}   \nonumber \\
 +  &  \frac{\varepsilon_0}{2}\sum_{n=1}^\infty\sum_{m=1}^{2n}(-1)^{n+m}\varepsilon_{2n}\partial_t^{m}\bm{E\cdot}\partial_t^{2n-m}\bm{\nabla\times E}   \nonumber \\ 
+  & \frac{\kappa_0}{2}  \sum_{n=1}^\infty\sum_{m=1}^{2n}(-1)^{n+m}\kappa_{2n}\partial_t^{m-1}\bm{B\cdot}\partial_t^{2n-m+1}\bm{\nabla\times B},  \label{rhomed}  \\
\bm{S}_\chi=&\frac{1}{2}\bm{D\times}\partial_t\bm{E}+\frac{1}{2}\bm{H\times}\partial_t\bm{B}.   \label{Smed}
\end{align}
It is straightforward to show that (\ref{rhomed}) and (\ref{Smed}) obey the conservation law (\ref{con}) when the macroscopic Maxwell equations (\ref{macmax1}) and (\ref{macmax2}) are used.

As in the cases of energy, momentum and angular momentum~\cite{phi11,phi12}, the expression for the zilch density (\ref{rhomed}) in a dispersive medium simplifies considerably for a time-averaged monochromatic beam. We consider the monochromatic $\bm{E}$ field
\begin{equation}  \label{Emono}
\bm{E}(\bm{r},t)=\frac{1}{2}\left(\bm{E}_0 (\bm{r})e^{-i\omega t}+\text{c.c}\right), 
\end{equation}
and a $\bm{B}$ field of the same form. Inserting these into  (\ref{rhomed}) and (\ref{Smed}) and taking a time average we obtain the monochromatic time-averaged zilch density $\bar{\rho}_\chi$ and flux $\bar{\bm{S}}_\chi$:
\begin{align}
\bar{\rho}_\chi=&\frac{\varepsilon_0\varepsilon(\omega)\omega}{2}\text{Im}\left(\bm{E}_0\bm{\cdot}\bm{B}^*_0\right)    \nonumber \\
& + \frac{\varepsilon_0\omega}{4}\frac{d\varepsilon(\omega)}{d\omega}\text{Re}\left(\bm{E}_0\bm{\cdot}\bm{\nabla\times E}^*_0\right)  \nonumber \\
 & - \frac{\kappa_0\omega}{4}\frac{d\kappa(\omega)}{d\omega}\text{Re}\left(\bm{B}_0\bm{\cdot}\bm{\nabla\times B}^*_0\right),  \label{rhomono0}  \\
\bar{\bm{S}}_\chi=&-\frac{\varepsilon_0\varepsilon(\omega)\omega}{4}\text{Im}\left(\bm{E}_0\bm{\times E}^*_0\right)       \nonumber \\  
&-\frac{\kappa_0\kappa(\omega)\omega}{4}\text{Im}\left(\bm{B}_0\bm{\times B}^*_0\right)    .  \label{Smono0}
\end{align}
The dispersive contributions to the time-averaged monochromatic zilch density (\ref{rhomono0}) contain first derivatives of the dielectric functions at the frequency of the beam; similar dispersive contributions occur in the momentum density~\cite{phi11}, angular momentum density~\cite{phi12}, and energy density (Eq.\ (\ref{brill}) below). The curls in (\ref{rhomono0}) can be rewritten using the Maxwell equations $\bm{\nabla\times E}_0=i\omega\bm{B}_0$, $\kappa(\omega)\bm{\nabla\times B}_0=-i\omega\varepsilon(\omega)\bm{E}_0/c^2$ for the monochromatic complex amplitudes. Utilizing also the definitions
\begin{equation} \label{ngnp}
n_g(\omega)=\frac{d[\omega n_p(\omega)]}{d\omega}, \qquad n_p(\omega)=\sqrt{\varepsilon(\omega)\mu(\omega)},
\end{equation}
of the group index $n_g(\omega)$ and the phase index $n_p(\omega)$ at the frequency of the wave, we find a simple formula for the time-averaged monochromatic zilch density (\ref{rhomono0}):
\begin{align} 
\bar{\rho}_\chi=&\frac{\varepsilon_0\omega}{2}n_g(\omega)\sqrt{\frac{\varepsilon(\omega)}{\mu(\omega)}}\text{Im}\left(\bm{E}_0\bm{\cdot}\bm{B}^*_0\right),     \label{rhomono}  \\
\bar{\bm{S}}_\chi=&-\frac{\varepsilon_0\omega}{4}\left[\varepsilon(\omega)\text{Im}\left(\bm{E}_0\bm{\times E}^*_0\right)     
+\frac{c^2}{\mu(\omega)}\text{Im}\left(\bm{B}_0\bm{\times B}^*_0\right)  \right]  ,  \label{Smono}
\end{align}
where we restate the time-averaged monochromatic zilch flux (\ref{Smono0}).

We apply the monochromatic results to a circularly polarized plane wave propagating in the $x$~direction. The complex amplitudes $\bm{E}_0$ and $\bm{B}_0$ of the electric and magnetic fields in such a beam are~\cite{jac}
\begin{gather}
\bm{E}_0 (\bm{r})=\mathcal{E}(\bm{e}_y\pm i\bm{e}_z)e^{ikx}, \quad \bm{B}_0 (\bm{r})=\mp\frac{i}{c}n_p(\omega)\bm{E}_0 (\bm{r}),    \label{EBcir}   \\
k=\frac{\omega}{c}n_p(\omega),  \qquad n_p(\omega)=\sqrt{\varepsilon(\omega)\mu(\omega)},
\end{gather}
where $\bm{e}_y$ ($\bm{e}_z$) are unit vectors in the $y$ ($z$)~directions, $k$ is the wave number, and $n_p(\omega)$ is the phase index as in (\ref{ngnp}). The time-averaged zilch density and flux of this wave are, from (\ref{rhomono}) and (\ref{Smono}),
\begin{equation}
\bar{\rho}_\chi= \pm\frac{\omega}{c} \varepsilon_0\varepsilon(\omega) n_g(\omega)\mathcal{E}^2 , \quad
\bar{\bm{S}}_\chi= \pm \omega \varepsilon_0\varepsilon(\omega)\mathcal{E}^2 \bm{e}_x.   \label{rhoScir}
\end{equation}
Dividing $\bar{\bm{S}}_\chi$ by $\bar{\rho}_\chi$ in (\ref{rhoScir}) we find that the optical zilch flows through the medium at the group velocity $c/n_g(\omega)$, just like the optical energy. To find the time-averaged zilch per unit energy in the circularly polarized plane wave, we require the time-averaged energy density $\bar{\rho}$ of a monochromatic beam in a dispersive medium, which is given by Brillouin's expression~\cite{jac,phi11}
\begin{equation}  \label{brill}
\bar{\rho}=\frac{\varepsilon_0}{4}\frac{d[\omega\varepsilon(\omega)]}{d \omega}\bm{E}_0\cdot\bm{E}_0^*+\frac{\mu_0}{4}\frac{d[\omega\mu(\omega)]}{d \omega}\bm{H}_0\cdot\bm{H}_0^*.
\end{equation}
For the plane wave (\ref{EBcir}), the time-averaged energy density (\ref{brill}) is
\begin{equation}  \label{encir}
\bar{\rho}= \frac{\varepsilon_0  n_p(\omega) n_g(\omega)}{\mu(\omega)}\mathcal{E}^2.
\end{equation}
Dividing $\bar{\rho}_\chi$ in (\ref{rhoScir}) by (\ref{encir}) we find the zilch per unit energy:
\begin{equation}  \label{chipere}
\frac{\bar{\rho}_\chi}{\bar{\rho}}= \pm\frac{\omega}{c} n_p(\omega)=\pm k.
\end{equation}
The zilch per unit energy of a circularly polarized plane wave is thus proportional to the phase index of the medium. This is in sharp contrast to the conserved spin angular momentum per unit energy of a circularly polarized monochromatic beam, which is independent of material properties even in a dispersive medium~\cite{phi12}. On the other hand, the  conserved linear momentum per unit energy of a plane wave is also proportional to the refractive index in a dispersive medium~\cite{phi12}. 

The zilch eigenstates of Sec.~\ref{sec:eigen} are easily generalized to a dispersive medium; circularly polarized plane waves are again eigenstates but the eigenvalues now acquire a factor of the refractive index:
\begin{equation}  
\chi=\pm\frac{\omega^2}{c}n_p(\omega).
\end{equation}

\section{Optical helicity in a dispersive medium}   \label{sec:hel}
A topic of recent discussion is whether the notion of optical chirality is best defined through zilch or through helicity~\cite{afa96,bli11,and12,col12,bar12,cam12}. While zilch and helicity are different quantities with different dimensions, they are proportional to each other for monochromatic beams in vacuum (and only for monochromatic beams). It is interesting to extend the comparison of zilch and helicity to a homogeneous, dispersive medium. This requires the expressions for optical helicity density and flux in the dispersive regime considered in the previous section. 

Optical helicity has its natural expression in terms of magnetic and electric vector potentials~\cite{des76,afa96}. As long as the medium is homogeneous we have $\bm{\nabla\cdot E}(\bm{r},t)=0$ and so we can introduce an ``electric" vector potential $\bm{C}(\bm{r},t)$ as follows:
\begin{equation}  \label{Cdef}
\bm{E}=-c\bm{\nabla\times C}.
\end{equation}
Maxwell's equations in the regime considered in the previous section then give the following relations between the electric (magnetic) field and the magnetic (electric) vector potential:
\begin{equation}  \label{EABC}
\bm{E}=-\partial_t\bm{A}, \qquad \kappa(i\partial_t)\bm{B}=-\frac{1}{c}\varepsilon(i\partial_t)\partial_t\bm{C},
\end{equation}
We first note that it is easy to generalize optical helicity density and flux to a non-dispersive medium where the dielectric functions are constants (rather than operators) in the time domain. The expressions are obtained from the vacuum helicity density and flux~\cite{des76,afa96} by inserting appropriate factors of the non-dispersive $\varepsilon$ and $\kappa(=1/\mu)$ (here we use SI units throughout):
\begin{align} 
\rho_h=&\frac{\varepsilon_0}{2}\left(c\kappa\bm{A\cdot B}-\varepsilon\bm{C\cdot E}\right),   \label{rhohnd}   \\
\bm{S}_h=&\frac{c\varepsilon_0}{2}\kappa\left(\bm{E\times A}+c\bm{B\times C}\right).  \label{Shnd} 
\end{align}
These expression were used in~\cite{bli12b} for a monochromatic beam. It is easy to show that (\ref{rhohnd}) and (\ref{Shnd}) obey the conservation law
\begin{equation}  \label{conh}
\partial_t\rho_h+\bm{\nabla\cdot S}_h=0
\end{equation}
because of the non-dispersive Maxwell equations (and relation between $\bm{E}$, $\bm{B}$ and the vector potentials). On the other hand, it is also easy to see that the conservation law (\ref{conh}) is not satisfied by (\ref{rhohnd}) and (\ref{Shnd}) if the medium is dispersive because $\varepsilon$ and $\kappa$ are then operators and the dispersive Maxwell equations do not give (\ref{conh}) no matter which vector fields the operators $\varepsilon(i\partial_t)$ and $\kappa(i\partial_t)$ are taken to act on in (\ref{rhohnd}) and (\ref{Shnd}). The situation here is no different than in the familiar case of electromagnetic energy~\cite{jac,phi11}: the energy density in a non-dispersive medium does not generalize in a simple way to a dispersive medium. There are non-trivial dispersive contributions to electromagnetic energy, momentum and angular momentum in media~\cite{jac,phi11,phi12}; we have seen that the same is true for zilch and now we must supply the dispersive contributions to helicity. The systematic way of finding the correct dispersive contributions is through Noether's theorem, as in Sec.~\ref{sec:med}. In the case of helicity however, the use of the standard electromagnetic action (\ref{act}) presents some difficulties~\cite{des76,cam12b,bli12}. Experience gained through solving the cases of energy-momentum, angular momentum and zilch allows us to simply write down the correct dispersive contributions to the helicity density. Whereas finding the generalization of a electromagnetic conserved quantity to a dispersive medium can be challenging, it is a straightforward matter to check whether any given generalization is correct, i.e. whether it satisfies the conservation law (\ref{conh}). Consider the expressions
\begin{align} 
\rho_h=& \frac{\varepsilon_0}{2} \Bigg\{  c\bm{A\cdot} \left[\kappa(i\partial_t)\bm{B}\right]-\left[\varepsilon(i\partial_t)\bm{C}\right]\bm{\cdot E}   \nonumber \\
 +  &  \sum_{n=1}^\infty\sum_{m=1}^{2n}(-1)^{n+m}\varepsilon_{2n}\partial_t^{m-1}\bm{C\cdot}\partial_t^{2n-m+1}\bm{E}   \nonumber \\ 
+  & c \left.  \sum_{n=1}^\infty\sum_{m=1}^{2n}(-1)^{n+m}\kappa_{2n}\partial_t^{m-1}\bm{A\cdot}\partial_t^{2n-m+1}\bm{B}  \right\},  \label{rhoh}   \\
\bm{S}_h=&\frac{c\varepsilon_0}{2}\left\{\bm{E\times}\left[\kappa(i\partial_t)\bm{A}\right]+c\left[\kappa(i\partial_t)\bm{B}\right]\bm{\times C}\right\}.  \label{Sh} 
\end{align}
These satisfy (\ref{conh}) with use of the macroscopic Maxwell equations and the relations (\ref{Cdef}) and (\ref{EABC}). The expressions (\ref{rhoh}) and (\ref{Sh}) also give vacuum helicity density and flux as a special case; they are therefore the correct generalization of conserved optical helicity to a homogeneous dispersive medium.

For a monochromatic wave (\ref{Emono}), the time-averaged helicity density and flux are easily found from (\ref{rhoh}) and (\ref{Sh}). The calculation is almost identical to that for zilch and we aid the comparison by eliminating the complex amplitudes of the vector potentials using 
\begin{equation}  \label{EABCmono}
\bm{A}_0=-\frac{i}{\omega}\bm{E}_0, \qquad \bm{C}_0=-i\frac{c\kappa(\omega)}{\omega\varepsilon(\omega)}\bm{B}_0,
\end{equation}
which follow from (\ref{EABC}). This gives the monochromatic time-averaged helicity density $\bar{\rho}_h$ and flux $\bar{\bm{S}}_h$ as
\begin{align} 
\bar{\rho}_h=&\frac{c\varepsilon_0n_g(\omega)}{2\omega\mu(\omega) n_p(\omega)}\text{Im}\left(\bm{E}_0\bm{\cdot}\bm{B}^*_0\right),     \label{rhohmono}  \\
\bar{\bm{S}}_h=&-\frac{c\varepsilon_0}{4\omega n_p^2(\omega)}\Bigg[\varepsilon(\omega)\text{Im}\left(\bm{E}_0\bm{\times E}^*_0\right)  \nonumber \\   
& \qquad\qquad\qquad + \left. \frac{c^2}{\mu(\omega)}\text{Im}\left(\bm{B}_0\bm{\times B}^*_0\right)  \right]  .  \label{Shmono}
\end{align}
Comparing these with the corresponding results (\ref{rhomono}) and (\ref{Smono}) for zilch, we find the following proportionality between helicity and zilch for any time-averaged monochromatic wave in a dispersive medium:
\begin{equation}  \label{chih}
\bar{\rho}_h=\frac{c}{\omega^2n_p^2(\omega)}\bar{\rho}_\chi,   \qquad\bar{\bm{S}}_h=\frac{c}{\omega^2n_p^2(\omega)}\bar{\bm{S}}_\chi.  
\end{equation}
This proportionality is familiar in the vacuum case $n_p(\omega)=1$; we have shown that it contains the square of the phase index  in a dispersive medium.

The zilch results (\ref{rhoScir}) for the circularly polarized plane wave (\ref{EBcir}), together with the proportionality (\ref{chih}), immediately give us the time-averaged helicity density and flux of a circularly polarized plane wave in a dispersive medium:
\begin{align}
\bar{\rho}_\chi= \pm\frac{\varepsilon_0 n_g(\omega)}{\omega\mu(\omega)}\mathcal{E}^2 , \qquad
\bar{\bm{S}}_\chi= \pm\frac{c\varepsilon_0 }{\omega\mu(\omega)}\mathcal{E}^2 \bm{e}_x.   \label{rhoShcir}
\end{align}
The proportionality (\ref{chih}) implies that, like zilch, helicity moves through the medium at the group velocity. The helicity per unit energy is given by $\bar{\rho}_\chi$ in (\ref{rhoShcir}) divided by (\ref{encir}):
\begin{equation}    \label{hpere}
\frac{\bar{\rho}_h}{\bar{\rho}}= \pm\frac{1}{\omega n_p(\omega)}.
\end{equation}
The helicity per unit energy of a circularly polarized plane wave is thus inversely proportional to the phase index $n_p(\omega)$, whereas the zilch per unit energy (\ref{chipere}) is proportional to $n_p(\omega)$. Although helicity has the dimensions of angular momentum, the (spin) angular momentum per unit energy of a circularly polarized monochromatic beam is $\pm1/\omega$ even in a dispersive medium~\cite{phi12} and so behaves quite differently to helicity per unit energy (\ref{hpere}).

\section{Conclusions}
The symmetry underlying conservation of Lipkin's zilch is a simple transformation of the vector potential (Eq.\ (\ref{trans})). Identification of the symmetry transformation has allowed us to prove that circularly polarized plane waves are optical zilch eigenstates. These plane-wave eigenstates, however, are not the most interesting possibility for light beams that carry optical zilch~\cite{hen10,hen12,sch12,ros12}. A straightforward generalization of the symmetry transformation gives the symmetries underlying an infinite class of electromagnetic conserved quantities identified by Morgan~\cite{mor64}. Optical zilch and optical helicity are both conserved in a homogeneous, dispersive medium, in frequency ranges where absorption is negligible. For time-averaged monochromatic waves, zilch and helicity differ by a factor that contains the square of the phase index of the material. The time-averaged zilch per unit energy of a circularly polarized plane wave in a dispersive medium is proportional to the phase index, whereas the helicity per unit energy is inversely proportional to the phase index. In contrast to zilch and helicity, the conserved (spin) angular momentum per unit energy of a circularly polarized monochromatic beam in a dispersive medium is independent of the dielectric functions. These results demonstrate further contrasts between zilch, helicity, and spin angular momentum.

\section*{Acknowledgements}
I thank E.\ Hendry for interesting discussions. Thanks are also due to a referee for pointing out reference~\cite{des76} and recommending the inclusion of the helicity results.


\end{document}